\documentclass[aps,twocolumn,nofootinbib,superscriptaddress,preprintnumbers,pra,10pt]{revtex4-1}

\usepackage{float}
\usepackage{arydshln}
\usepackage{colortbl}
\usepackage{amsmath,amssymb}
\usepackage{dsfont} 
\usepackage{hyperref}
\usepackage{graphicx}
\usepackage{enumitem}
\usepackage{arydshln}
\usepackage{mathtools}
\usepackage{bbold}
\usepackage{soul}
\usepackage{slashed}
\usepackage{amsmath,bm}
\usepackage{yfonts}
\usepackage{lipsum}
\usepackage{slashed}
\usepackage{mathtools}
\usepackage{physics}

\makeatletter
\g@addto@macro\bfseries{\boldmath}
\makeatother

\newcommand{\be} {\begin{equation}}
\newcommand{\ee} {\end{equation}}
\newcommand{\bea} {\begin{eqnarray}}
\newcommand{\eea} {\end{eqnarray}}
\newcommand{\no} {\nonumber}

\newcommand{\cL}{{\mathcal L}}
\newcommand{\cJ}{{\mathcal J}}
\newcommand{\cA}{{\mathcal A}}
\newcommand{\cB}{{\mathcal B}} 
 
\renewcommand{\Re}{{\rm Re}}

\begin{document}

\preprint{ZU-TH-36/22}
\title{Semi-inclusive Lepton Flavor Universality ratio in $b\to s\ell^+\ell^-$ transitions}

\author{Marco Ardu}
\affiliation{LUPM, CNRS, Université Montpellier Place Eugene Bataillon, F-34095 Montpellier, Cedex 5, France}
\author{Gino Isidori }
\author{Marko Pesut}

\affiliation{Physik-Institut, Universit\"at Zu\"rich, CH-8057 Z\"urich, Switzerland}

\begin{abstract}
\vspace{5mm}
We construct a semi-inclusive Lepton Flavour Universality (LFU) ratio, $R_\Sigma$, to test  $\mu/e$ universality in $b\to s\ell^+\ell^-$ transitions
at $e^+e^-$ $B$-meson factories. Combining different decay channels, this observable maximises the sensitivity to possible LFU violations 
of short distance origin, yet preserving a clean theoretical interpretation in case of a deviation 
from its Standard Model prediction, $R_\Sigma^{\rm SM}=1$. 
\vspace{3mm}
\end{abstract}

\maketitle
\allowdisplaybreaks

\section{Introduction}
Within the Standard Model (SM) all the lepton Yukawa couplings are small compared to the SM gauge couplings, 
giving rise to an approximate accidental symmetry known as  Lepton Flavour Universality (LFU) (see e.g.~Ref.~\cite{Artuso:2022ijh}).
In the last few years precise LFU tests have been  performed by the LHCb experiment
in rare $B$-meson decays. More precisely,  $\mu/e$ universality has been tested via measurements of the 
exclusive ratios~\cite{Hiller:2003js}
\be
R_M  = 
\frac{\Gamma[B \to M  \mu^+ \mu^-]}{\Gamma[B \to M e^+ e^-]},
\label{eq:RK}
\ee
in specific dilepton invariant mass intervals, and for different final state mesons ($M=K^+, K_S,  K^{*0}, K^{*+}$).

Within the SM, ${R_M^{\textrm{SM}} =1}$ up to corrections due to phase space and QED, 
which do not exceed $1\%$\cite{Bordone:2016gaq,Isidori:2020acz,Isidori:2022bzw} for the observables 
so far considered.\footnote{The only exception is $R_{K^*}$ for  $2 m_\mu \leq m_{\ell\ell} \leq 1$~GeV, 
whose SM prediction is~$0.906\pm 0.028$~\cite{Bordone:2016gaq}.}
 The experimental results 
reported by LHCb are all below this figure~\cite{LHCb:2021trn,LHCb:2017avl,LHCb:2021lvy}
and, if combined, even in the most conservative way~\cite{Isidori:2021vtc}, provide a strong evidence of 
physics beyond the SM. Given the potential ground-breaking impact of this result, it would be 
extremely important to confirm it in different experimental conditions. 

An ideal setup for completely independent tests of $\mu/e$ universality in rare $B$ decays 
is provided by experiments performed at $e^+e^-$ $B$-meson factories, such as  Belle-II~\cite{Belle-II:2018jsg}.
These benefit of a much cleaner environment, compared to experiments at hadron colliders, such as LHCb.
The only serious drawback, at least in the short term, is the limited statistics.
To overcome this limitation, we propose in this letter to test the same short-distance 
dynamics via a semi-inclusive LFU ratio,
\be
R_\Sigma  = 
\frac{ \sum_{H_s} \Gamma[B \to H_s \mu^+ \mu^-]}{ \sum_{H_s}  \Gamma[B \to H_s e^+ e^-]}.
\label{eq:RSig}
\ee
Here $B$ stands for $B$ mesons of any charge (i.e.~$B^\pm$ and $B^0$), while 
$H_s$ denotes a series of well-defined exclusive final states, and
an appropriate kinematical projection to maximise the statistics and, 
at the same time, retain a clean 
sensitivity to possible LFU effects of short-distance origin. 

The main ideas behind the construction of $R_\Sigma$ can be listed as follows:
\begin{itemize}
\item{}
A possible violation of $\mu/e$ universality of 
short-distance origin, i.e.~a violation attributed to a local $b\to s \ell^+\ell^-$ interaction, should manifest 
in any exclusive $B \to H_s \ell^+ \ell^-$ decay. We can therefore combine many exclusive channels 
to increase the statistics. The key point to address is how to combine the different channels, taking into 
account the uncertainties due to unknown hadronic matrix elements.
\item{}
In principle, the simplest solution would be to consider a fully inclusive final state $|X_s\rangle$, of strangeness $|S|=1$. 
However, this is quite challenging from the experimental point of view, requiring an independent (opposite-side)
$B$-meson tag that usually implies a low efficiency. On the other hand, as pointed in~\cite{Isidori:2021tzd}, 
LFU ratios have a rather constrained structure that allow us to combine them even in absence 
of a complete description of the underlying hadronic dynamics. Following the approach of  Ref.~\cite{Isidori:2021tzd},
we can therefore limit the combination only to the specific sum of final states which have an easy (self-tagged) signature.
The only strict requirement is to select the same combination of hadrons, in the 
same kinematical range,  for both $\ell=e$ and $\ell=\mu$.
\item{}
Following the above prescription, within the SM we expect $R^{\rm SM}_\Sigma=1$ up to QED corrections.
Less obvious is how to interpret the result if $R^{\rm SM}_\Sigma\not =1$. To this purpose, the key observation is that 
in a large fraction of the phase space the $b\to s \ell^+\ell^-$  SM amplitude is dominated by the product of a 
left-handed hadronic current times an (almost) left-handed leptonic current~\cite{Hiller:2014ula}.
Hence the LFU ratios project out the left-handed component of the possible non-SM amplitude. 
The only exception is the region of narrow charmonia resonances, and the low-$m_{\ell\ell}$ region, which are 
dominated by lepton-universal contributions. Cutting out the latter with appropriate kinematical cuts we can
build a semi-inclusive ratio that maximises the sensitivity and allow for a clean theoretical interpretation.
\end{itemize}
Taking into account the above considerations, we proceed with the detailed definition of $R_\Sigma$.

\section{Definition of $R_\Sigma$}

\subsection{Hadronic state and kinematical range}
The hadronic states we propose to analyse together are composed by an odd number of kaons 
and an arbitrary number of pions. The set can be limited to charged pions and
kaons only, but could also include a few neutral states (which notoriously have smaller
detection efficiencies). We can generically denote the set as
\be
|H_s \rangle \in \{  (2n+1) |  K\rangle +   m | \pi  \rangle    \}\,, \qquad n,m \in \mathbb{N}\,.
\label{eq:Hs}
\ee
As anticipated, it is essential to ensure the same hadronic composition 
(i.e.~the same $n$ and $m$, and the same number of $\pi^0$ and $K_S$)  
for  both $\ell=e$ and $\ell=\mu$. On the other hand, neutral and charged $B$-meson
decays can be combined in the semi-inclusive sum.

In order to define the kinematical range for the dilepton invariant mass, two requirements need to be fulfilled:
i)~avoiding the region of the narrow charmonia, $J/\Psi$ and $\Psi(2S)$, which would dilute 
a possible LFU-violating effect of short-distance origin, ii)~performing kinematical cuts that do not 
induce LFU-violating effects of QED origin (i.e.~$\alpha_{\rm em} \log m_\ell$ corrections).

As demonstrated in Ref.~\cite{Isidori:2020acz},  the dangerous QED collinear logs are avoided if the dilepton range is defined in terms of the 
collinear-safe variable
\be
q_0^2 = (p_B - p_H)^2\,,
\ee
where $p_H$ is the sum of all hadronic momenta. Contrary to experiments performed at hadron colliders, 
the variable $q_0^2$, which coincides with $m^2_{\ell\ell}$ only in the limit of negligible final-state radiation,
is accessible at $e^+e^-$ $B$ factories. 
Defining cuts in $m^2_{\ell\ell}$, 
rather than in $q_0^2$, is the main reason why 
the estimates of the inclusive ratios presented in~\cite{Huber:2015sra,Huber:2020vup} (including QED corrections) are significantly 
different for electrons and muons.

\begin{figure}[t]
\centering  
\includegraphics[scale = 0.48]{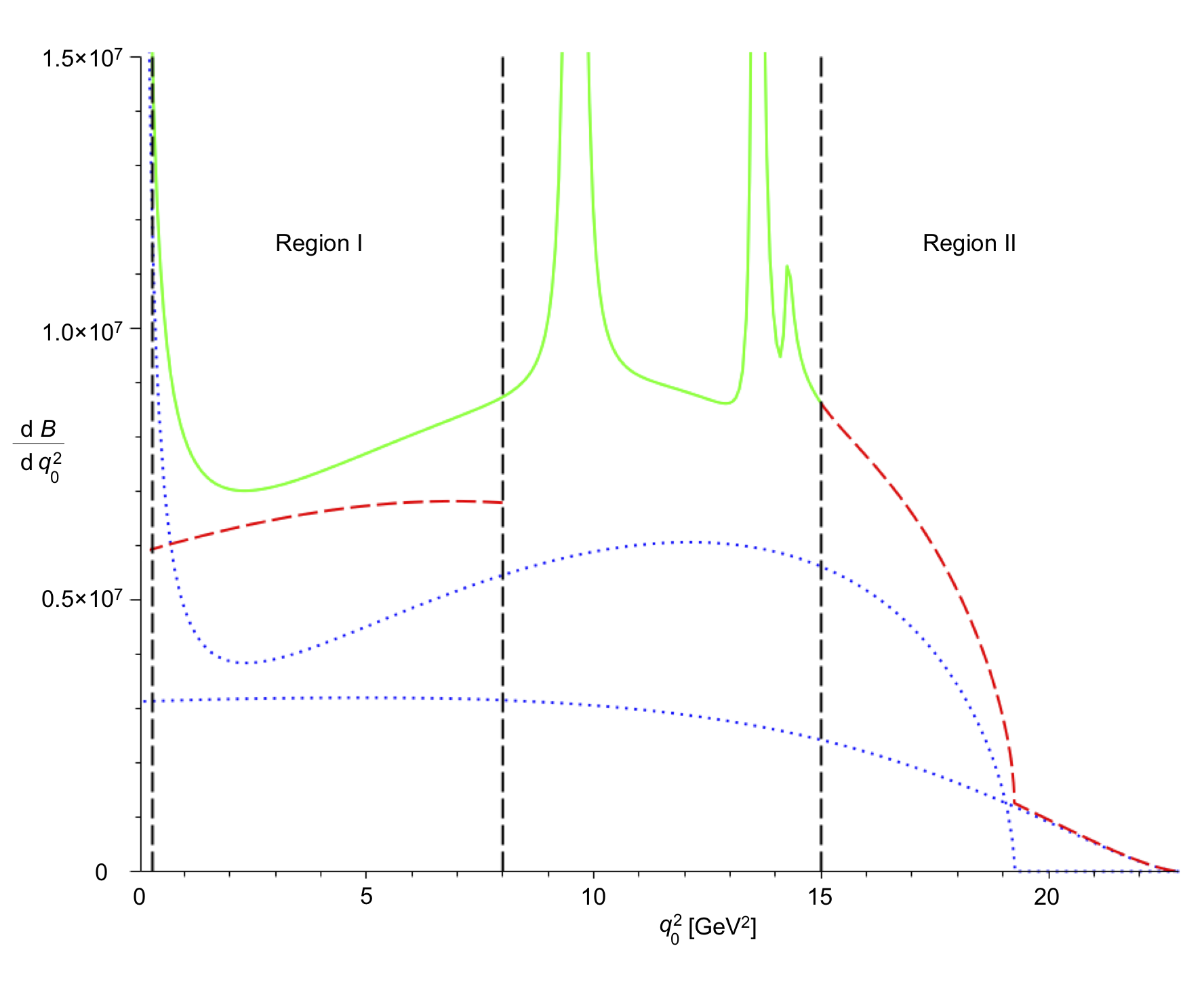}
\caption{Semi-inclusive dilepton spectrum obtained by summing 
 $\cB(B \to K e^+ e^-)$  and $\cB(B\to K^*e^+ e^-)$. The (green) full line corresponds 
 to the complete sum, including an estimate of long-distance contributions. The (blue) dotted lines 
indicate the short-distance contributions for the two separate modes. The  (red) dashed line indicates 
the sum, after applying the longitudinal projection on the $B\to K^*e^+ e^-$ mode. The vertical dashed lines 
denote the two regions for the evaluation of $R_\Sigma$.
\label{Fig:spectum} }
\end{figure}

We recall that we are interested only in the LFU ratio,  and not in a precise description 
of the absolute decay probability in terms of short-distance dynamics. Hence we can 
afford to include a small (universal) long-distance contamination due to resonance 
tails in $R_\Sigma$.  Keeping this in mind, we can extend the low-$q_0^2$ window  up to $8~{\rm GeV}^2$,
which is safely below the $J/\Psi$ peak, and define the high-$q_0^2$ window starting from
$15~{\rm GeV}^2$.  In order to avoid the Dalitz decays ($P \to \ell^+\ell^-\gamma$)
of light mesons, it is also useful to set a lower cut $q_0^2 \geq q_{\rm min}^2 = 0.3~{\rm GeV}^2 \approx m_\eta^2$.
Summarising, as shown in Fig.~\ref{Fig:spectum}, we propose to define $R_\Sigma$ integrating over the following two $q_0^2$ windows:
\bea
&{\rm Region\ I:}  & \quad 0.3~{\rm GeV}^2 \leq q_0^2 \leq 8~{\rm GeV}^2\,,  \no\\
&{\rm Region\ II:} & \quad q_0^2 \geq 15~{\rm GeV}^2\,.
\label{eq:q0range}
\eea

The $R_\Sigma$ defined in Eq.~(\ref{eq:RSig}), with $|H_s \rangle$ in (\ref{eq:Hs}) and $q^2_0$ in~(\ref{eq:q0range}), is a good variable 
to test LFU: it satisfies
\be
R_\Sigma^{\rm SM}=1.00 \pm 0.01\,,
\ee
where the error is due to subleading QED corrections, and is dominated by short-distance dynamics. However, its interpretation 
if $R^{\rm exp}_\Sigma \not =1$ would not be very clean, the main problem being the photon-pole contribution in Region~I: 
a lepton-universal amplitude which necessarily dilutes a possible LFU effect.  
 The precise estimate of this dilution requires the knowledge of the hadronic matrix elements of the dipole operator $Q_7$ 
 (see Appendix~\ref{sect:App}), which are unknown for multi-meson final states. The effect can be suppressed 
reducing further the $q_0^2$ range, and partially dealt with by  treating the unknown hadronic matrix elements as nuisance parameters~\cite{Isidori:2021tzd}. 
However, as we discuss below, a more efficient strategy in case of $B$-factory experiments is to 
get rid of  the photon-pole contribution with a simple angular projection.

\subsection{Angular projection in the low--$q_0^2$ region}
\label{sect:Phat}

In order to define the angular projection which allow us to get rid of  the photon-pole contribution,
it is worth to discuss first the allowed values of the 
angular momentum of the hadronic system ($J_{\rm had}$), and the  helicity structure of the decay amplitude.

\paragraph{Allowed values of $J_{\rm had}$.} The matrix elements
of the $b\to s\ell^+\ell^-$ effective Lagrangian  (see Appendix~\ref{sect:App}), which are non-vanishing at the tree level in 
 $B\to H_s \ell^+\ell^-$ decays, can be generally decomposed as
\be
    \bra{H_s} \cJ_{\rm had}^\mu \ket{B}  \times \cJ_{\rm lept}^\mu\,.
\ee
Since  the $B$ meson has vanishing angular momentum ($J^{\rm P}  =
0^-$) and  $\cJ_{\rm had}^\mu$ transforms as a Lorentz vector, 
this implies that $J_{\rm had}=0$ or $1$. In principle, higher $J_{\rm had}$ can be generated by truly non-local contributions of the
four-quark operators (via multipole expansion); however, these effects are extremely suppressed. For all practical purposes we can 
restrict the attention to $J_{\rm had}=0$ and $J_{\rm had}=1$. 

\paragraph{Helicity structure of the decay amplitude.}
The photon-pole contribution in a given $B\to H_s \ell^+ \ell^-$ decay is present if the $B\to H_s \gamma$ transition, with an on-shell photon, is allowed.
If $J_{\rm had}=0$, or if $J_{\rm had}=1$ and is longitudinally polarized ($J_{\rm had}=1^0$), helicity conservation 
forbids the $B\to H_s \gamma$ decay.
Isolating the $J_{\rm had}=0$ and $J_{\rm had}=1^0$ partial waves in the decay rate of $B\to X_s \ell \ell$ would thus permit to neglect the photon-pole contribution.

\paragraph{Definition of the projection.}
The helicity of the dilepton system (and, correspondingly, of the hadronic one) can be identified experimentally via the the angle $\theta_\ell$, defined as the angle between the lepton 
and the $B$ direction of flight in the $q_0$ rest frame.\footnote{The $q_0$ rest frame coincides with dilepton  rest frame in the limit of negligible final-state radiation~\cite{Isidori:2020acz}.}
Neglecting lepton masses, the $B\to H_s \ell^+ \ell^-$ double-differential decay rate can be decomposed as~\cite{Kruger:2005ep} 
\bea
  \frac{d^2\Gamma}{dq_0^2\ d\cos\theta_\ell}  &=& \cA^{(1)}_{[J=0,1^0]} \times  \sin^2\theta_\ell +  \cA^{(2)}_{[J=1^\pm]} \times  \cos\theta_\ell   \no\\
&+&  \cA^{(3)}_{[J=1^\pm]}  \times (1+\cos^2\theta_\ell)~. 
 \label{eq:Adec}
 \eea
 As explicitly indicated, the photon-pole contribution can show up only in the $\cA^{(2)}$ and $\cA^{(3)}$ coefficients. 
 We can easily get rid of these terms  with a non-trivial integration over  $x\equiv \cos \theta_\ell$, or better acting with 
 the following projection operator
\begin{equation}
   \hat{P} \left[  \frac{d^2\Gamma}{dq_0^2\ dx} \right]  = \int_{-1}^{+1}  dx \, (2-5 x^2)  \frac{d^2\Gamma}{dq_0^2\ dx} \,.
   \label{eq:Phat}
\end{equation}
The projector is such that
\bea
&&    \hat{P}(1+x^2)=\hat{P}(x)=0\,,  \\
&&  \hat{P}(1-x^2)= \frac{4}{3} = \int_{-1}^{+1} dx\, (1-x^2)\,.
\label{eq:Pnorm}
\eea
The normalization of the projector in Eq.~(\ref{eq:Pnorm}) is such that  $\hat{P}$ acts like the identity operator on 
the $J_{\rm had}=0$ and $J_{\rm had}=1^0$ components of the decay rate. 

 A similar projection procedure has been introduced 
 in Ref.~\cite{Huber:2015sra}, in the context 
 of fully inclusive $B\to X_s \ell^+\ell^-$ decays,
 assuming the SM effective Lagrangian.
 We stress that the decomposition in Eq.~(\ref{eq:Adec}),
 and the projection operator in Eq.~(\ref{eq:Phat}), holds for any 
 final state $|H_s\rangle$ and also in presence of (local) new-physics contributions.

We further stress that there is no need to operate with $\hat{P}$ neither in the case of a single kaon in the final state ($|H_s\rangle = |K^\pm\rangle$ or $|K_S\rangle$), 
which necessarily has $J_{\rm had}=0$, nor in the high-$q^2_0$ region (Region II), where the photon-pole contribution is already 
strongly suppressed by the value of $q_0^2$. The longitudinal projector $\hat{P}$ has to be applied only on multi-meson final states 
in Region I. As shown in Fig.~\ref{Fig:spectum}, where we illustrate the impact of the projection
on the exclusive decay $B\to K^* (\to K \pi) \ell^+\ell^-$, the loss of statistics due to the projection 
is quite limited (below $20\%$).

\section{$R_\Sigma$ beyond the SM}

Following Ref.~\cite{Isidori:2021tzd},  the explicit expression of  $d\hat\Gamma^\ell_{H_s}/dq_0^2$  in terms of Wilson coefficients
in generic extensions of the SM can be written as 
\bea
 \frac{d\hat\Gamma^{\ell}_{H_s}}{dq_0^2}  &=& f_{H_s}^{\ell}(q^2) \Big\{ \left( \left|C^{\ell}_L\right|^2 + \left| C^{\ell\prime}_L\right|^2 
+  \Re\Big[  \eta^0_H(q^2)  C^{\ell*}_L C^{\ell\prime}_L  \Big]\right)
   \no \\  && \qquad  + ( L \to R) +O(C_7) \Big\}\,.
\label{eq:rx_dGdq}
\eea
The $O(C_7)$ terms indicate contributions of the dipole operator  which are not enhanced by the photon pole, or arise by 
the  high-$q^2_0$ region where the longitudinal projector is not applied. Given the  smallness of $|C_7|$ in the SM~\cite{Blake:2016olu},
the experimental bounds on non-standard 
$b\to s$ dipole transitions~\cite{LHCb:2020dof},
and given that  the dipole amplitude is lepton-flavor universal, they can be safely neglected.

In the (well-justified) limit of neglecting the $O(C_7)$ terms, $
R_\Sigma$ assumes the following simple from
\be
R_\Sigma = \frac{   \Big\{ \left|C^{\mu}_L\right|^2 + \left| C^{\mu\prime}_L\right|^2  + \Re\left[  \langle \eta^0_{\Sigma} \rangle  C^{\mu*}_L C^{\mu\prime}_L  \right] \Big\} + \Big\{ L\to R \Big\} 
   }{      \Big\{ \left|C^{e}_L\right|^2 + \left| C^{e\prime}_L\right|^2  + \Re\left[  \langle \eta^0_{\Sigma} \rangle  C^{e*}_L C^{e\prime}_L  \right] \Big\} + \Big\{ L \to R \Big\} }\, .
\label{eq:RXfull}
\ee
The expression depends on a single combination of hadronic parameters,
$\langle \eta^0_{\Sigma} \rangle$, which controls the  relative weight of 
vector and axial currents in the semi-inclusive sum
(averaged over the various hadronic states and over the 
different $q_0^2$ regions).  As noted in 
Ref.~\cite{Isidori:2021tzd}, the positivity of the 
decay rate implies 
$|\langle \eta^0_{\Sigma} \rangle|\leq 2$. 
In the exclusive $B\to K$ transition 
$\langle \eta^0_{K} \rangle =2$,  while 
$|\langle \eta^0_{\Sigma} \rangle| \approx 0$
if the sum over $H_s$ is sufficiently inclusive.

To check how close we get to a sufficiently inclusive sum, considering  only few hadronic states, we have analysed numerically the case where 
the $|H_s\rangle$ set is limited to $|K\rangle$ and $|K^*\rangle$ 
(whose hadronic form factors are known). The semi-inlcusive 
dilepton spectrum thus obtained is shown in 
Fig.~\ref{Fig:spectum}. 
The numerical analysis has been performed employing the 
$B\to K$ and $B\to K^*$ form factors from Ref.~\cite{Ball:2004ye,Ball:2004rg}.
For illustrative purposes, in Fig.~\ref{Fig:spectum} 
we also included the effect of the narrow charmonia 
states in the $B\to K \ell^+\ell^-$ case, following the procedure developed in~\cite{Cornella:2020aoq}.

Applying the projection operator only in 
the $B\to K^* \ell^+\ell^-$ case, and only in Region~I, we estimate $R_\Sigma$ allowing 
for non-universal contributions to the Wilson coefficients.
Defining
\be
\Delta C_i = \Delta C_i^\mu - \Delta C_i^e~,
\ee
and expanding for small $|\Delta C_i|$, we can write
\bea
R_\Sigma - 1  &=&   \kappa_{L} \Delta C_L + \kappa_{R} \Delta C_R \no \\
&+& \kappa_{L^\prime} \Delta C_L^\prime + \kappa_{R^\prime} \Delta C_{R}^\prime + O(\Delta C_i^2)~.
\eea
The numerical coefficients obtained with the procedure outlined above are 
\bea
& \kappa_{L}= 0.25\pm0.02\,, \quad \qquad  &|\kappa_{L^\prime}| < 0.01\,,   \\
& \kappa_{R} =-0.02\pm0.03\,, \ \qquad  &|\kappa_{R^\prime}| < 0.01\,.
\eea
The errors are dominated by the uncertainty on the SM Wilson
coefficients. In particular, we include a conservative $20\%$
error on the value of $C_9^{\rm SM}$, to account for the uncertainties associated to non-local contributions from four-fermion operators
(see e.g.~\cite{Ciuchini:2015qxb,Arbey:2018ics,Gubernari:2020eft}).

The smallness of the $\kappa_{L,R}^\prime$ coefficients 
indicates that we are already very close to the
inclusive limit 
(i.e.~$\langle \eta^0_{\Sigma} \rangle \approx 0$)
even when we consider only $|K\rangle$ and $|K^*\rangle$ states. 
We thus conclude that $R_\Sigma$ is a clean and sensitive 
probe of a single combination of LFU-violating
Wilson coefficients:
\be
\Delta C_L = (C_{9}^{\mu}-C_{9}^{e})-(C_{10}^{\mu}-C_{10}^{e})\,.
\ee
Interestingly enough, a non-vanishing $\Delta C_L$ 
is usually advocated, 
both for phenomenological and 
model-building considerations, 
as the origin of the violations 
of universality so far observed in 
$b\to s\ell^+\ell^-$ transitions
(see e.g.~\cite{Buttazzo:2017ixm,Alguero:2018nvb,Altmannshofer:2021qrr,Cornella:2021sby}).

\section{Conclusions}
The semi-inclusive LFU ratio $R_\Sigma$, 
defined by  Eqs.~(\ref{eq:RSig}),  (\ref{eq:Hs}) and (\ref{eq:q0range}), 
 possibly improved by the angular projection discussed in Sect.~\ref{sect:Phat}, could allow for very clean testing of LFU even with limited statistics.
The semi-inclusive transitions we propose to analyse,
summing events in the two $q_0^2$ regions, and taking into account the angular projection, correspond
to an effective branching ratio $\cB_{\rm eff} \approx 
2 \times 10^{-6}$
(the precise value depends on how many exclusive channels 
will be included). Taking into account that both 
charged and neutral $B$-meson decays can be combined, 
this is about 30 times the statistics available to measure 
$R_{K^+}$ in the low-$m_{\ell\ell}$ region, as defined 
in the LHCb analysis~\cite{LHCb:2021trn}.

The Belle-II experiment has already observed the $B \to K^* \ell^+\ell^-$ transition~\cite{Belle-II:2022fky}, and the statistics collected since then has more than doubled. 
With such statistics, $R_\Sigma$ could possibly be measured 
with a O($10\%$) error, providing an interesting non-trivial  
independent test of $\mu/e$ universality in $b\to s\ell^+\ell^-$ transitions.

%%%%%%%%%%%%%%%%%%%%%%%%%%%%%%%%%%%%%%%%%%%%%%%%%%
\section*{Acknowledgements}
%%%%%%%%%%%%%%%%%%%%%%%%%%%%%%%%%%%%%%%%%%%%%%%%%%
We thank Marzia Bordone, Patrick Owen, and Nicola Serra for useful discussions.
This project has received funding from the European Research Council (ERC) under the European Union's Horizon 2020 research and innovation programme under grant agreement 833280 (FLAY), and by the Swiss National Science Foundation (SNF) under contract 200020\_204428. M.A. is supported by a doctoral fellowship from the IN2P3 and thanks the Physik-Institut of the University of Zurich for its hospitality during the completion of this work.

\appendix
\section{$b\to s\ell^+\ell^-$ effective Lagrangian}
\label{sect:App}
At energies below the electroweak scale, short-distance effects of both the SM and heavy New Physics can be described by contact interactions among the light SM fields. The effective Lagrangian relevant for the $b\to s\ell^+ \ell^-$ transitions that we are interested in is
\begin{equation}
	\cL^{b\to s \ell \ell}_{\rm eff}= -2\sqrt{2} G_F \frac{\alpha_e}{4\pi}V^*_{ts}V_{tb}\sum_i C_i \mathcal{O}_i + \text{h.c}
\end{equation} 
where
\begin{align}
	\mathcal{O}_7=\frac{m_b}{e}(\overline{s}_L\sigma_{\mu\nu} b_R)F^{\mu\nu},\qquad \mathcal{O}'_7=\frac{m_b}{e}(\overline{s}_R\sigma_{\mu\nu} b_L)F^{\mu\nu},  \nonumber\\
	\mathcal{O}^\ell_{9}=(\overline{s}_L\gamma_{\mu} b_L)(\overline{\ell} \gamma^{\mu}\ell),\qquad 	\mathcal{O}^\ell_{10}=(\overline{s}_L\gamma_{\mu} b_L)(\overline{\ell} \gamma^{\mu}\gamma_5\ell),  \nonumber\\
	\mathcal{O}^{\ell\prime}_{9}=(\overline{s}_R\gamma_{\mu} b_R)(\overline{\ell} \gamma^{\mu}\ell),\qquad 	\mathcal{O}^{\ell\prime}_{10}=(\overline{s}_R\gamma_{\mu} b_R)(\overline{\ell} \gamma^{\mu}\gamma_5\ell).
\end{align}
We do not consider scalar operators because their chirally-suppressed contribution to $b\to s\ell^+ \ell^-$ amplitudes are irrelevant for the observables considered in the ratio $R_\Sigma$ defined in Eq.~(\ref{eq:RSig}). Moreover, as discussed in the text, the non-local lepton-universal effects of four-quark operators are taken into account via an effective shift 
(and corresponding uncertainty) 
in the SM value of $C_9$. 

Taking advantage of the (almost) left-handed structure of SM lepton currents, NP effects can be best distinguished re-writing the operator basis in terms of chirally projected operators for the leptons
\begin{align}
	C_L^\ell=C_9^\ell-C_{10}^\ell\qquad C_L^{\ell\prime}=C_9^{\ell\prime}-C_{10}^{\ell\prime} \nonumber\\
	C_R^\ell=C_9^\ell+C_{10}^\ell\qquad C_R^{\ell\prime}=C_9^{\ell\prime}+C_{10}^{\ell\prime}
\end{align} 
with the additional advantage that, in rates, the $L,R$ interference  is suppressed by the small lepton masses. The decay amplitude of a generic $B\to H_s \ell^+ \ell^-$ process is decomposed as
\begin{align}
	\mathcal{A}^\ell(B\to H_s \ell^+ \ell^-)\propto (\mathcal{M}^\alpha_{H_s,L})^\ell (\cJ^L_\alpha)^\ell+L\leftrightarrow R
\end{align} 
where $(\cJ^X_\alpha)^\ell=\overline{\ell}_X\gamma_\alpha \ell_X$, with $X=L,R$, and 
\begin{align}
    (\mathcal{M}^\alpha_{H_s,X})^\ell=C^\ell_X \cJ^\alpha_{H_s}+C^{\ell\prime}_X \cJ^{\prime\alpha}_{H_s}+C_7 \cJ^{7\alpha}_{H_s}
\end{align}
with
\begin{align}
	\cJ^\alpha_{H_s}&=\bra{H_s}(\overline{s}_L\gamma^{\alpha} b_L)\ket{B},\qquad\cJ^{\prime\alpha}_{H_s}=\bra{H_s}(\overline{s}_R\gamma^{\alpha} b_R)\ket{B},\nonumber\\
	\cJ^{7\alpha}_{H_s}&\propto \frac{q_\beta}{q^2} \bra{H_s}(\overline{s}_L\sigma^{\alpha\beta} b_R)\ket{B},
 \label{eq:Hsamp}
\end{align}
where $q$ is the four-momentum of the lepton pair.

\bibliography{main}
\end{document}